# OPTIMA: Design-Space Exploration of Discharge-Based In-SRAM Computing: Quantifying Energy-Accuracy Trade-Offs

*Saeed Seyedfaraji, †Severin Ja¨ger, *Salar Shakibhamedan, *Asad Aftab, ‡Semeen Rehman

*,†,‡Technische Universita¨t Wien (TU Wien), Vienna, Austria
‡University of Amsterdam, The Netherlands

*{saeed.seyedfaraji, salar.shakibhamedan, asad.aftab}@tuwien.ac.at, †e1613004@student.tuwien.ac.at, ‡s.rehman@uva.nl

*Abstract*—In-SRAM computing promises energy efficiency, but circuit nonlinearities and PVT variations pose major challenges in designing robust accelerators. To address this, we introduce OPTIMA, a modeling framework that aids in analyzing bit-line discharge and power consumption in 6T-SRAM-based accelerators. It provides insights into limiting factors and enables fast design-space exploration of circuit configurations. Leveraging OPTIMA for in-SRAM multiplications exhibits ∼100× simulation speed-up while maintaining an RMS modeling error of 0.88 mV. Exploration yields an optimized multiplier with 1.05 pJ energy consumption per 4-bit operation and classification accuracies of 71.8 % (top-1) and 90.4 % (top-5) for ImageNet and 92.5 % for CIFAR-10 datasets respectively when applied in quantized DNNs. To further support research and development, we made our tool flow available open source at https://github.com/sevjaeg/optima.

## I. INTRODUCTION

Over recent decades, semiconductor scaling has significantly advanced computing performance. Despite the exponential growth in transistors, modern processors face limitations in power consumption, operating frequency, and single-thread performance [1]. The memory wall [2] and the power wall [3] constrain applications not by processor throughput but by energy consumption and memory system performance. Although complex memory hierarchies mask these issues, workloads like Machine Learning (ML) and big data suffer from high memory latencies, causing processor idle time and performance degradation [4].

In systems-on-chip, power is not only consumed by the computational logic, but also in the memory system, for example in Static Random Access Memory (SRAM) leakage power and off-chip interconnects for Dynamic Random Access Memory (DRAM). Data movement between memory and computation units dominates, accounting for over 60% of the total energy use [5]. New computing paradigms based on emerging non-volatile memory or vertical memory integration

The authors gratefully acknowledge funding from European Union's Horizon 2020 Research and Innovation programme under the Marie Skłodowska Curie grant agreement No. 956090 (APROPOS: Approximate Computing for Power and Energy Optimisation, http://www.apropos.eu/).

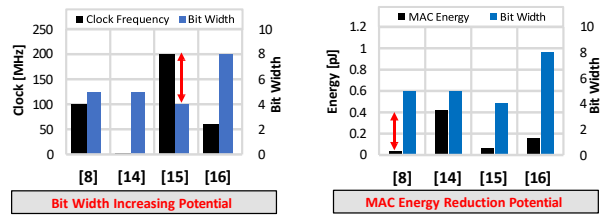

Fig. 1: State-of-the-art in-SRAM multiplication design space

show potential in reducing energy per data transfer or off-chip communication energy [6], [7]. However, minimizing data movement requires innovative paradigms like Near-Memory Computing (NMC) or In-Memory Computation (IMC). NMC adds computational cores close to the memory system, while IMC enhances the memory itself with computational capabilities. This minimizes data movement and yields optimal energy efficiency. This work specifically explores IMC.

IMC targets energy-efficient computation by reducing memory transfer time and energy for tasks like ML inference and operations such as Multiplication and Accumulation (MAC) and matrix-vector multiplication. Leveraging the existing cache area of modern Central Processing Units (CPUs) for SRAM-based IMC can enhance energy efficiency without significant architectural changes. Prior to incorporating IMC methodologies, it is crucial to initially discern the attributes of the diverse alternatives. Within the array of memory technologies employed for IMC, 6T SRAM presents notable sub-bank divisibility (enabling parallel computing), convenient accessibility, and economical fabrication costs [8], [9]. Hence, 6T SRAM becomes a promising candidate for IMC.

In-SRAM computing has two major flavors: bit-line (BL) computing and discharge-based computing. The former relies on concurrent activation of word lines (WLs) and logic primitives in readout amplifiers, often augmented with digital NMC logic. This approach accelerates various applications, including search [10], floating-point arithmetic [11], and neural networks [12]. On the other hand, discharge-based computing utilizes the analog nature of read-out transistors to create a data-dependent discharge on SRAM's BLs. This allows operations like addition [13] or multiplication [8] directly in the memory array, promising even greater energy savings.

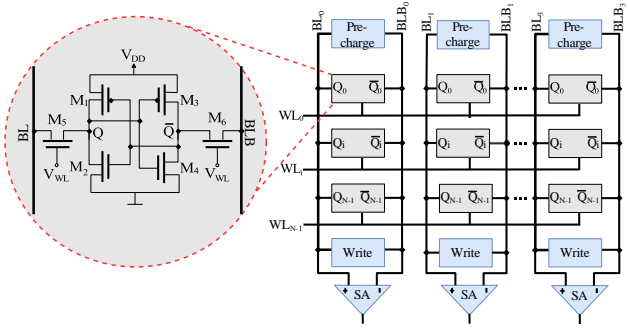

Fig. 2: 6T SRAM cell and SRAM array

Thus, this work emphasizes discharge-based SRAM IMC.

## A. Problem Motivation and Research Challenge

Discharge-based in-SRAM computing circuits are typically optimized for energy consumption, latency, accuracy, or area resulting in a rich design space. Fig. 1 compares different design points for IMC multipliers [8], [14]–[16]. Clearly, there are significant trade-offs between the design metrics. For instance, the circuit presented in [16] shows higher bit widths thus, offering high accuracy, whereas the other circuits demonstrated higher latency thus, they are significantly slower compared to the multiplier demonstrated in [15].

These trade-offs have to be investigated thoroughly with design-space exploration to find (Pareto-)optimal configurations. However, the analog nature of these circuits leads to large configuration spaces which are typically evaluated with slow circuit simulations based on solving differential equations. In addition, Process, Voltage, and Temperature (PVT) variations have to be considered, adding significant run-time overhead. Therefore, the selection of an optimal analog-based IMC configuration is very time-consuming, thus inefficient, and needs to be addressed.

## B. Our Contributions

To address the above-discussed research challenge, we propose a modeling technique called OPTIMA. This manuscript's contributions are as follows:

- In the first step, we identify and analyze the error sources at the circuit level and assess their impact on SRAM IMC circuits in Section III.
- Subsequently, in Section IV we develop a precise model for SRAM BL discharge and power consumption, considering nonlinearities and PVT variations. To validate our model, we utilize circuit simulation data from a 65 nm technology transistor model.
- In Section V, we showcase OPTIMA by performing fast design-space exploration of a 4-bit multiplication circuit. We derive a circuit configuration with an optimized energy-accuracy tradeoff.
- Finally, we assess the effectiveness of the optimized in-memory multiplier in Deep Neural Networks (DNNs). In Section VI, we evaluate the accuracy with four standard networks on two major data sets.

## II. BACKGROUND

### A. Static Random-Access Memory (SRAM)

Fig. 2 illustrates the standard six-transistor (6T) SRAM cell consisting of two cross-coupled inverters ($M_1$ & $M_2$, $M_3$ & $M_4$) and two access transistors ($M_5$, $M_6$). The data bit is differentially stored at nodes $Q$ and $\overline{Q}$. To connect a memory cell to the differential BL and Bit-Line-Bar (BLB) (which are shared among multiple cells in an array), the WL voltage $V_{WL}$ is employed. In typical computing systems, memory cells are organized in arrays, as exemplified in Fig. 2, depicting an array of N words, each consisting of four SRAM cells. A standard SRAM cell has two basic operations:

- *Read*: Firstly, both BLs are pre-charged to the supply voltage $V_{DD}$. Then, the address decoder activates the cell by driving the WL to $V_{DD}$. As either $Q$ or $\overline{Q}$ is 0 V, one of the BLs is discharged via $M_2$ or $M_4$ while the complementary BL remains at $V_{DD}$. This differential output signal is captured by Sense Amplifiers (SAs).
- *Write*: Updating the data stored in the memory cell starts with pre-charging both BLs to $V_{DD}$. Then, the BLs are driven to the desired voltages by discharging one of them to 0 V via the write circuitry. Finally, activating the corresponding WL lets the BL voltage propagate to the memory cell. Given proper transistor dimensions, the BL overwrites the cell data with the desired write data.

The operational principle of SRAM is heavily dependent on analog effects, which opens up creative possibilities for using the circuits discussed in this section for in-memory operations.

### B. Discharge-Based IMC

This paradigm relies on operating SRAM cells off-spec and is also referred to as current-domain in-SRAM computing [17]. The fundamental idea involves storing one operand in the memory cell while applying the other operand via the WL. As shown in Fig. 3, this can be used to implement multiplications. Firstly, a weight (in this case, '1') is stored in the memory cell, and the BLs are pre-charged to $V_{DD}$. Next, the input data is applied to the WL. If the weight $d$ is '0', the BL remains at $V_{DD}$. However, if it is '1', the WL voltage results in a discharge on BLB which is proportional to the product of the operands:

$$\delta V(t) \propto V_{WL} \cdot d \cdot t \qquad (1)$$

In the simplest case, the WL voltage is binary so that readout can be achieved with adapted SAs. However, this principle can be extended to multiple-bit computations with two key ideas:

1. Quantizing the WL voltage with a Digital-to-Analog Converter (DAC) [8]. The discharge is proportional to this voltage and captured with an Analog-to-Digital Converter (ADC) after a constant sampling time to obtain a product of the cell's data and the applied WL voltage.

2. Interpreting the data of multiple memory cells as a bit vector representing an N-bit integer. This requires a mechanism to represent bit weights which can be achieved in the time, charge, or current domain with reasonable overhead [18]. The weighted discharges are then combined, and the overall discharge is sampled with an ADC.

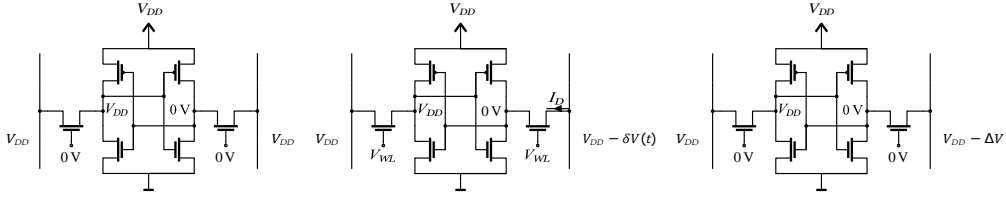

Fig. 3: Multiplication sequence: holding '1' with pre-charged BLs (left), discharge with driven WLs (center), final result (right)

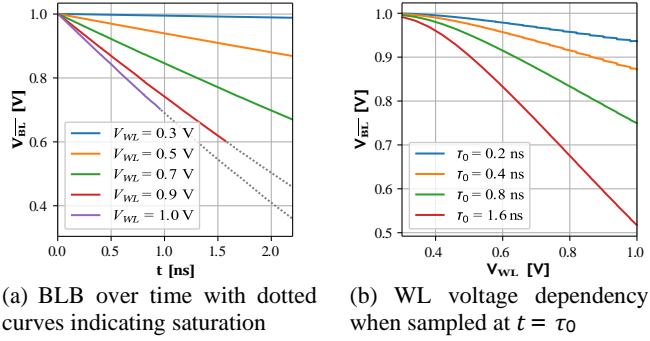

(a) BLB over time with dotted curves indicating saturation
(b) WL voltage dependency when sampled at $t = \tau_0$

Fig. 4: BLB discharge non-idealities.

A synthesis of these ideas enables multiplications of two multi-bit words [8]. However, these in-SRAM multiplications are currently limited to small bit widths (1-5 bits).

### III. IN-SRAM COMPUTING ERROR SOURCES

To understand why practical implementations of in-SRAM multiplication circuits are limited to small bit widths, we analyze the effects of non-idealities and operating condition variations at the circuit level. Even though the error sources are presented independently in this section, they all co-occur in any discharge-based IMC circuit. The data in this section is based on circuit simulation for a TSMC 65 nm Complementary Metal-Oxide Semiconductor (CMOS) technology. Further error sources with subordinate impact are analyzed in [19].

*1) Circuit Nonlinearity:* In-6T-SRAM multipliers operate based on a data-dependent current through the pass transistor. When the stored data is zero, there is no discharge, and the BLB voltage remains $V_{DD}$. However, if the data is '1' and a voltage representing '0' is applied via the WL, a small discharge occurs due to the non-zero source-drain current of the Metal Oxide Semiconductor Field Effect Transistor (MOSFET) at $V_{th}$ (see Fig. 4a). This asymmetry can lead to different results for $a \times b$ and $b \times a$, degrading the multiplier's accuracy.

Moreover, the quadratic current-voltage relationship of MOSFETs introduces a nonlinear discharge dependency on the applied WL voltage (see Fig. 4b). The quantization performed with a conventional DAC consequently leads to nonlinear multiplication results. The adoption of a nonlinear DAC is a potential solution [15], even though its practical circuit implementation poses significant challenges.

Another non-ideality arises from the transition of pass transistors from the saturation to the linear region with increasing BLB discharges. The saturation condition for $M_6$ is:

$$V_{\overline{BL}} \geq V_{WL} - V_{th} \qquad (2)$$

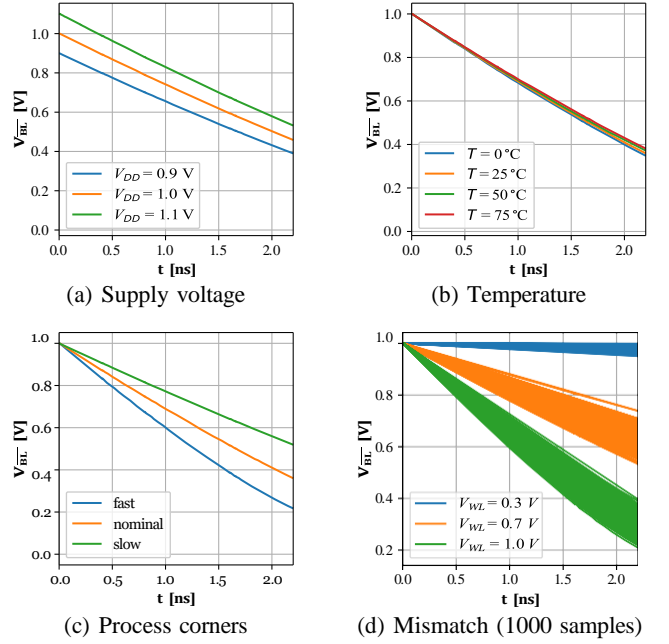

(a) Supply voltage
(b) Temperature
(c) Process corners
(d) Mismatch (1000 samples)

Fig. 5: Influence of PVT variations on the BLB discharge in TSMC 65 nm technology

If the BLB discharges below this threshold, the transistor enters the linear region, which leads to a reduced current and therefore slower discharge. Selecting an appropriate ADC sampling time $\tau_0$ ensures that the pass transistors remain in saturation during the BLB discharge. However, smaller $\tau_0$ values lead to reduced BLB voltage swings, degrading the Signal to Noise Ratio (SNR).

*2) PVT Variations:* PVT variations are the second major limiting factor for accurate in-6T-SRAM computing. Their effect on the discharge voltage is shown in Fig. 5. While temperature fluctuations vary the discharge speed only slightly, supply voltage and process variations alter the discharge voltage curves significantly. Note that supply voltage changes do not only affect the SRAM circuit, but also the thresholds of ADCs and DACs. The variations are also data dependent, e.g. in Fig. 5d, the transistor mismatch-induced deviation grows with the applied WL voltage.

### IV. OUR MODELING FRAMEWORK: OPTIMA

Our modeling framework OPTIMA is available open source[1] and facilitates the simulation of analog BL voltage in an event-based fashion, akin to digital simulation tools. This promises significantly shorter runtimes. We accomplished this through the following steps:

---
[1]The source code is available at https://github.com/sevjaeg/optima.

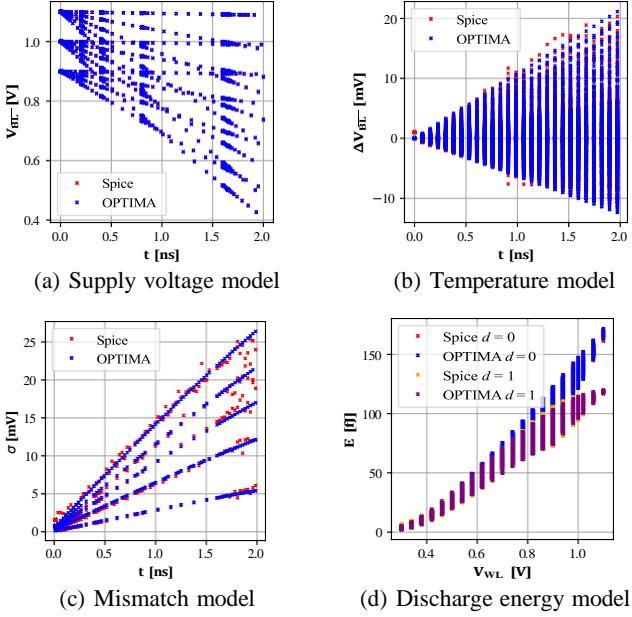

Fig. 6: OPTIMA discharge modeling evaluation

- Execute thorough multi-corner circuit simulations.
- Develop behavioral models for essential analog metrics, integrating non-idealities and accounting for PVT variations using circuit equations and simulation data.
- Incorporate these models into a versatile discrete-time simulation framework written in System Verilog. Our model is adaptable to a wide range of discharge-based in-SRAM operations.

The behavioral modeling within OPTIMA consists of parameterized discharge and energy models:

### A. Discharge Models

To model the BLB discharge accurately, OPTIMA follows an iterative approach. Firstly, the voltage is modeled as a function of time and WL voltage. In the next steps, variation sources are added. All models are based on polynomial functions. In this manuscript, $p_n(X)$ defines a polynomial of the variable $X$ of degree $n$ with $n+1$ coefficients.

As expressed in Eq. 1, the voltage on the BLB is time-dependent and influenced by the WL voltage. However, as illustrated in Fig. 4, this relationship is nonlinear. Therefore, the function

$$V_{\overline{BL}}(t, V_{WL}) = V_{DD} + p_4(V_{od}) \cdot p_2(t) \quad (3)$$

with the overdrive voltage $V_{od} = V_{WL} - V_{th}$ is used to model the BLB discharge. This model already incorporates the nonlinearities as presented in Section III-1.

This model is extended by a supply voltage function

$$V_{\overline{BL}}(t, V_{WL}, V_{DD}) = V_{\overline{BL}}(t, V_{WL}) \cdot p_2(\Delta V_{DD}) \quad (4)$$

with $\Delta V_{DD} = V_{DD} - V_{DD,\text{nom}}$.

As depicted in Fig. 5b, temperature has only a minor effect. Therefore, it is modeled as an additive error term for deviations from the nominal temperature $T_{\text{nom}}$ using the function

$$V_{\overline{BL}}(t, V_{WL}, V_{DD}, T) = V_{\overline{BL}}(t, V_{WL}, V_{DD}) \quad (5)$$
$$+ (t \cdot (T - T_{\text{nom}}) \cdot p_3(V_{WL}))$$

In contrast to the parameters discussed so far, process variations are an inherently stochastic property. Therefore, they are modeled with a statistical approach. As mismatch causes Gaussian variations in the BLB voltage, its standard deviations $\sigma$ are modeled as:

$$\sigma(t, V_{WL}) = p_3(t) \cdot p_3(V_{WL}) \quad (6)$$

### B. Energy Models

To assess energy-accuracy trade-offs in SRAM IMC, the energy consumption of the cells has to be modeled as well. It is mainly related to charging the BL capacitances during the pre-charge phases.

The energy consumed in writes is data-independent due to the symmetric cell layout and can thus be modeled as:

$$E_{\text{wr}}(V_{DD}, T) = p_2(V_{DD}) \cdot p_1(T) \quad (7)$$

In contrast, the discharge energy depends on the BLB discharge which depends on both operands (data $d$ and WL voltage):

$$E_{\text{dc}}(d, V_{DD}, V_{WL}, T) = p_1(V_{DD}) \cdot p_3(\Delta V_{\overline{BL}}) \cdot p_1(T) \quad (8)$$

The BLB discharge $\Delta V_{\overline{BL}}$ depends on $d$, $V_{DD}$, $V_{WL}$, and $T$ and is calculated with the models presented in Eq. 3–5.

### C. Model Evaluation

Least-squares fitting is employed to determine the coefficients for the models in Eq. 3–8 based on extensive simulation data. These parameters are subsequently incorporated into the discrete-time simulation model. For the transistor mismatch, the Gaussian distribution with $\sigma$ from Eq. 6 is sampled for each discharge. The resulting voltages and energies are illustrated in Fig. 6. The Root Mean Square (RMS) modeling errors are 0.76 mV (basic discharge), 0.88 mV (VDD), 0.76 mV (temperature), 0.59 mV (mismatch $\sigma$), 0.15 fJ (write energy), and 0.74 fJ (discharge energy) respectively. These values are below typical ADC Least Significant Bit (LSB) voltages for in-SRAM operations, indicating a sufficient accuracy for reliable analyses of in-SRAM computing circuits.

## V. CASE STUDY: IN-SRAM MULTIPLIER

We showcase the application of OPTIMA using the 4-bit multiplication circuit presented in [8]. This circuit employs the discharge principle illustrated in Fig. 3 applied to a 4-bit per word array, as depicted in Fig. 2. The WLs voltages are controlled using a 4-bit DAC, and the discharge occurs at intervals of $\tau_0$, $2\tau_0$, $4\tau_0$, and $8\tau_0$ on the different BLBs to implement bit weighing. Subsequently, the discharge voltages are sampled using switches and capacitors, and the combined discharge voltage is captured using an ADC. For simplicity, we omit the analog accumulation step from the original publication and concentrate on the multiplication process, which can be modeled efficiently with OPTIMA.

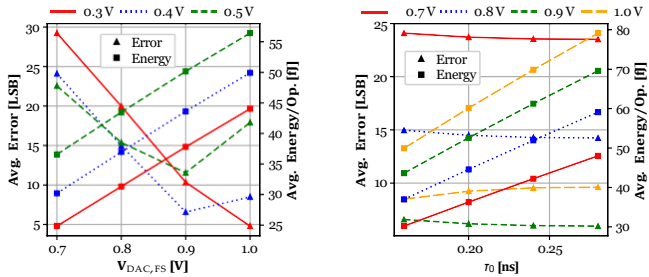

Fig. 7: Design space corners simulations with OPTIMA for different values of $V_{DAC,0}$ with $\tau_0 = 1.6\,\text{s}$ (left) and $V_{DAC,FS}$ with $V_{DAC,0} = 0.4\,\text{V}$ (right).

To demonstrate our methodology, we establish a design space characterized by three circuit parameters:

- $\tau_0$: Discharge time of the least significant BLB
- $V_{DAC,0}$: Output voltage of the DAC for data word '0'
- $V_{DAC,FS}$: Full-scale output voltage of the DAC

We select 48 design corners and simulate the circuit using the OPTIMA framework. The results of this design-space exploration are depicted in Fig. 7. A higher value of $V_{DAC,FS}$ results in a linear increase in energy consumption but is associated with higher accuracies in most cases. Increasing $V_{DAC,0}$ or $\tau_0$ also leads to higher energy consumption. The former has a positive impact on multiplication errors, while the latter has minimal influence on accuracy.

We select three interesting configurations to perform PVT analyses with OPTIMA. The first corner *fom* is selected based on maximizing a Figure of Merit (FOM) combining the averages of error after quantization $\epsilon_{mul}$ and energy per operation $E_{mul}$:

$$\text{FOM} = \frac{1}{\overline{\epsilon_{mul}} \cdot \overline{E_{mul}}} \qquad (9)$$

The second corner *power* is the one with the minimum energy per multiplication and the third corner *mismatch* shows the smallest standard deviation at the maximum discharge (i.e., it is least impacted by process variation). Table I summarizes the corresponding parameters.

The PVT analysis results, including sampling mismatch corners, are presented in Fig. 8. Deviations in average multiplication results signify the impact of circuit nonlinearities, while high standard deviations indicate susceptibility to mismatch. In terms of the *power* configuration, issues are observed in both cases, with the *variation* corner performing notably worse than *fom* for small values. However, for large values, it demonstrates robustness against process variation. Voltage and temperature fluctuations also exert a significant effect on the error level, with the *fom* corner proving to be the least susceptible to these variations.

Drawing conclusions from our design-space exploration using OPTIMA, we determine that the *fom* configuration produces the most favorable results. At an operating frequency of 167 MHz, it exhibits an average multiplication error of 4.8 LSBs. The worst-case analog standard deviation is 5.04 mV. For a single operation, including write and multiplication, the average energy consumption is 1.05 pJ. For the multiplication circuit, OPTIMA achieves a speedup of 10× for iteration over the input space and design corners and 28.1× for mismatch Monte Carlo (MC) sampling compared to circuit simulation in Cadence Virtuoso.

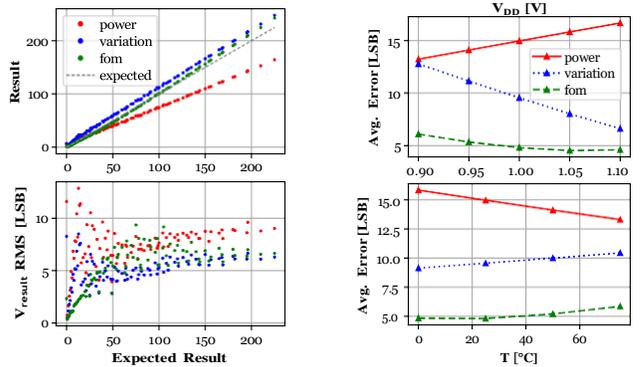

Fig. 8: Average multiplication results and analog standard deviations (left) as well as influence of voltage and temperature variations on the error (right) for the selected corners.

TABLE I: Selected design corners

| Corner | $\tau_0$ | $V_{DAC,0}$ | $V_{DAC,FS}$ | $\overline{\epsilon_{mul}}$ | $\overline{E_{mul}}$ |
|---|---|---|---|---|---|
| *fom* | 0.16 ns | 0.3 V | 1.0 V | 4.78 | 44 fJ |
| *power* | 0.16 ns | 0.3 V | 0.7 V | 15 | 37 fJ |
| *variation* | 0.24 ns | 0.4 V | 1.0 V | 9.6 | 69.8 fJ |

## VI. APPLICATION ANALYSIS

In this research, we assess the effectiveness of the proposed in-SRAM multiplier configurations for DNN inference. To accomplish this, we use these multipliers in deep learning models for image classification, specifically VGG16, VGG19 [20], ResNet50, and ResNet101 [21], which are trained on the ImageNet dataset [22]. Furthermore, we apply these models to the CIFAR-10 [23] dataset to assess their performance.

In the initial series of experiments, pre-trained DNNs obtained from the Keras model zoo, originally trained on the ImageNet dataset, are employed to showcase the efficacy of the selected in-memory multiplier configurations. The pre-trained DNNs utilize a FLOAT32 number representation. These experiments aim to utilize our proposed multiplier for all multiplication operations within these DNNs. To achieve this, we quantize the pre-trained DNNs to an INT4 number representation using post-training quantization.

This quantization process adheres to the specifications of TensorFlow Lite, with INT8 being replaced by INT4 along with corresponding adjustments to the range, restrictions, and other specifications. Subsequently, retraining procedures are implemented to mitigate the impact of quantization on the relevant metrics. The proposed in-memory multiplier configurations are then applied to execute all multiplication operations within the DNNs.

Given the application's emphasis on image classification, the evaluation focuses on top-1 and top-5 accuracies as key metrics. The results, depicting the utilization of the in-memory multiplier in three configurations for the quantized ImageNet-trained DNNs, are presented in Table II.

TABLE II: DNN classification accuracies for ImageNet

| Model | Number of Multiplications [×10^9] | Baseline FLOAT32 | | Baseline INT4 | | In-Memory *fom* (INT4) | | In-Memory *power* (INT4) | | In-Memory *variation* (INT4) | |
|---|---|---|---|---|---|---|---|---|---|---|---|
| | | Top-1 Accuracy [%] | Top-5 Accuracy [%] | Top-1 Accuracy [%] | Top-5 Accuracy [%] | Top-1 Accuracy [%] | Top-5 Accuracy [%] | Top-1 Accuracy [%] | Top-5 Accuracy [%] | Top-1 Accuracy [%] | Top-5 Accuracy [%] |
| VGG16 | 15.61 | 70.30 | 90.10 | 69.25 | 89.62 | 68.97 | 89.11 | 64.45 | 81.79 | 38.22 | 47.81 |
| VGG19 | 19.77 | 71.30 | 90.00 | 70.09 | 89.78 | 69.91 | 89.24 | 63.34 | 79.61 | 36.66 | 48.37 |
| ResNet50 | 4.14 | 74.90 | 92.10 | 73.48 | 91.75 | 73.39 | 91.65 | 61.56 | 80.88 | 48.07 | 56.71 |
| ResNet101 | 7.87 | 76.40 | 92.80 | 75.12 | 91.91 | 74.95 | 91.63 | 59.77 | 78.49 | 48.45 | 53.19 |

TABLE III: DNN classification accuracies for CIFAR10

| Model | FLOAT32 Top-1 Accuracy [%] | INT4 Top-1 Accuracy [%] | In-Memory *fom* (INT4) Top-1 Accuracy [%] | In-Memory *power* (INT4) Top-1 Accuracy [%] | In-Memory *variation* (INT4) Top-1 Accuracy [%] |
|---|---|---|---|---|---|
| VGG16 | 92.24 | 92.04 | 91.98 | 87.39 | 68.10 |
| VGG19 | 92.71 | 92.42 | 92.29 | 89.79 | 66.85 |
| ResNet50 | 93.10 | 92.86 | 92.83 | 90.81 | 73.83 |
| ResNet101 | 93.35 | 93.06 | 93.04 | 90.42 | 69.77 |

To further evaluate our findings, a second set of experiments is conducted on DNNs to classify the CIFAR-10 dataset. Again, the DNNs is obtained by employing INT4 quantization. Additionally, the last layer is replaced with a fully-connected layer containing 10 neurons (reflecting the number of classes in CIFAR-10), and transfer learning is utilized for training. The same experimental conditions as before are applied to assess the performance of the multiplier configurations. The multiplication operations in each DNN closely aligned with the values in Table II, as the modifications were confined to the last layer, resulting in minimal changes ($< 0.03\%$) in the number of multiplication for an individual inference.

Among the selected multiplier configurations, the *fom* corner is most effective in our application. It exhibits a marginal decrease in accuracy compared to quantized INT4 DNNs (considered as baseline) and outperforms all other configurations across all DNNs. For the ImageNet dataset, the top-1 accuracies degrade by a mere 1.42% and 0.18% and for the CIFAR-10 dataset, by 0.315% and 0.06% compared to FLOAT32 and INT4 representations. The *power* configuration trades off accuracy for reduced energy consumption and shows more significant accuracy reductions. Interestingly, the *variation* corner achieves only top-1 accuracies of 42.85% and 69.63% for ImageNet and CIFAR-10, respectively on average. The reason is its high error level for multiplications with small operands (see Fig. 8), which dominate in DNN workloads.

## VII. CONCLUSION

To overcome the challenges of selecting one design for in-SRAM computing circuits, we have introduced OPTIMA. It is a fast yet accurate design-space exploration technique exhibiting ∼100 simulation speed-up while maintaining an RMS modeling error of 0.88 mV for an in-SRAM multiplier. Our techniques' exploration yields an optimized multiplier with 1.05 pJ energy consumption per 4-bit operation, and classification accuracies of 71.8% (top-1) and 90.4% (top-5) for ImageNet and 92.5% for CIFAR-10 datasets, respectively, when applied in quantized DNNs. We made our tool flow available to the community at https://github.com/sevjaeg/optima.

## REFERENCES

[1] K. Rupp. 50 years of microprocessor trend data, 2022.
[2] W. A. Wulf and S. A. McKee. Hitting the memory wall. *ACM SIGARCH Computer Architecture News*, 23(1):20–24, 1995.
[3] M. Horowitz. 1.1 Computing's energy problem (and what we can do about it). In *2014 IEEE International Solid-State Circuits Conference Digest of Technical Papers (ISSCC)*, pp. 10–14, 2014.
[4] M. Ferdman *et al.* Clearing the clouds: A study of emerging scale-out workloads on modern hardware. In *Proceedings of the Seventeenth International Conference on Architectural Support for Programming Languages and Operating Systems*, ASPLOS XVII, pp. 37–48, 2012.
[5] A. Boroumand *et al.* Google workloads for consumer devices: Mitigating data movement bottlenecks. *SIGPLAN Not.*, 53(2):316–331, 2018.
[6] A. Gebregiorgis *et al.* A survey on memory-centric computer architectures. *ACM Journal on Emerging Technologies in Computing Systems (JETC)*, 18(4):1–50, 2022.
[7] Y. Yu and N. K. Jha. Energy-efficient monolithic three-dimensional on-chip memory architectures. *IEEE Transactions on Nanotechnology*, 17(4):620–633, 2017.
[8] M. Ali *et al.* IMAC: In-memory multi-bit multiplication and ACcumulation in 6T SRAM array. *IEEE Transactions on Circuits and Systems I: Regular Papers*, 67(8):2521–2531, 2020.
[9] S. Seyedfaraji *et al.* SMART: Investigating the impact of threshold voltage suppression in an in-SRAM multiplication/accumulation accelerator for accuracy improvement in 65 nm CMOS technology. In *2022 25th Euromicro Conference on Digital System Design (DSD)*, pp. 821–826. IEEE, 2022.
[10] S. Aga *et al.* Compute caches. In *2017 IEEE International Symposium on High Performance Computer Architecture (HPCA)*, pp. 481–492, 2017.
[11] J. Wang *et al.* A 28-nm compute SRAM with bit-serial logic/arithmetic operations for programmable in-memory vector computing. *IEEE Journal of Solid-State Circuits*, 55(1):76–86, 2020.
[12] C. Eckert *et al.* Neural cache: Bit-serial in-cache acceleration of deep neural networks. In *2018 ACM/IEEE 45th Annual International Symposium on Computer Architecture (ISCA)*. IEEE, 2018.
[13] M. Kang *et al.* A multi-functional in-memory inference processor using a standard 6T SRAM array. *IEEE Journal of Solid-State Circuits*, 53(2):642–655, 2018.
[14] K. Sanni *et al.* A charge-based architecture for energy-efficient vector-vector multiplication in 65nm CMOS. In *2018 IEEE International Symposium on Circuits and Systems (ISCAS)*, pp. 1–5. IEEE, 2018.
[15] S. Seyedfaraji *et al.* AID: Accuracy improvement of analog discharge-based in-SRAM multiplication accelerator. In *2022 Design, Automation & Test in Europe Conference & Exhibition (DATE)*. IEEE, 2022.
[16] M. Gong *et al.* A 65nm thermometer-encoded time/charge-based compute-in-memory neural network accelerator at 0.735 pJ/MAC and 0.41 pJ/update. *IEEE Transactions on Circuits and Systems II: Express Briefs*, 68(4):1408–1412, 2020.
[17] Z. Chen *et al.* CAP-RAM: A charge-domain in-memory computing 6T-SRAM for accurate and precision-programmable CNN inference. *IEEE Journal of Solid-State Circuits*, 56(6):1924–1935, 2021.
[18] Z. Lin *et al.* A review on SRAM-based computing in-memory: Circuits, functions, and applications. *Journal of Semiconductors*, 43(3):031401, 2022.
[19] A. Kneip and D. Bol. Impact of analog non-idealities on the design space of 6T-SRAM current-domain dot-product operators for in-memory computing. *IEEE Transactions on Circuits and Systems I: Regular Papers*, 68(5):1931–1944, 2021.
[20] K. Simonyan and A. Zisserman. Very deep convolutional networks for large-scale image recognition. *arXiv preprint arXiv:1409.1556*, 2014.
[21] K. He *et al.* Deep residual learning for image recognition. In *Proceedings of the IEEE conference on computer vision and pattern recognition*, pp. 770–778, 2016.
[22] J. Deng *et al.* Imagenet: A large-scale hierarchical image database. In *2009 IEEE Conference on Computer Vision and Pattern Recognition*, pp. 248–255, 2009.
[23] A. Krizhevsky. Learning multiple layers of features from tiny images. *University of Toronto*, 2012.